# IMPACT OF THE TEMPERATURE AND HUMIDITY VARIATIONS ON LINK QUALITY OF XM1000 MOTE SENSORS

Samia Allaoua Chelloug

Networks and Communication Systems Department, College of Computer and Information Sciences, Princess Nourah bint AbdulRahman University, Riyadh Kingdom of Saudi Arabia

*ABSTRACT*

*The core motivations of deploying a sensor network for a specific application come from the autonomy of sensors, their reduced size, and their capabilities for computing and communicating in a short range. However, many challenges for sensor networks still exist: minimizing energy consumptions, and ensuring the performance of communication that may be affected by many parameters. The work described in this paper covers mainly the analysis of the impact of the temperature and humidity variations on link quality of XM1000 operating under TinyOS. Two-way ANOVA test has been applied and the obtained results show that both the temperature and humidity variations impact RSSI.*

## 1. INTRODUCTION

By definition, a sensor network is deployed either inside the phenomenon to be monitored or very close to it. Each sensor consists of a microcontroller, a transceiver, a source of power, and a sensing unit [1]. Sensor networks are being considered for use in: smart homes, E-health, industrial automation,… Some applications have stringent requirements that concern the delay and the reliability of communication [2]. However, recent studies show that signals fluctuate and sometimes it is very hard to ensure a certain degree of quality of service. Hence, the aim of our experimental study is to show how the temperature and humidity fluctuations may affect the link quality of XM1000 sensor motes that have been manufactured by ADVENTICS and use an MSP430 microcontroller. It provides a program flash memory and data program storage space. XM1000 functions under TinyOS2.x operating system which is appropriate for programming embedded systems. In particular, our experimental study focuses on understating the effect of the two parameters (temperature and humidity) upon link quality.

This article provides a literature review of related works. Sections 3 and 4 explains the hardware architecture of XM1000 and TinyOs. Sections 5 outlines the principles of ANOVA. Then, our experimental results are illustrated and analyzed by two-way ANOVA.

                                                                            21

International Journal of Ad hoc, Sensor & Ubiquitous Computing (IJASUC) Vol.5, No.6, December 2014

## 2. IMPACT OF METEROLOGICAL CONDITIONS ON SENSOR NETWORKS

Designing appropriate protocols for sensor networks is a critical task due to their hardware constraints. Many research works exploit some indicators that can be obtained via hardware to process and communicate data in wireless sensor networks. In particular, the research community faces the challenge of power control and power saving to extend the lifetime of wireless sensor networks. Power control and power saving may be applied jointly with other sensor protocols to efficiently localize sensors, route data packets, ovoid interference,… Extensive studies are based on measuring the strength of the received signal (RSSI) or the link quality (LQI) that can be calculated by the transceiver. Some results show that LQI is a good indicator for new sensor platforms [3]. Unfortunately, meteorological conditions (temperature, humidity) may affect the link quality. Hence, it is necessary to analyze the impact of such parameters to assure accuracy while controlling or saving power. Some research works provided experimental studies for understating the impact of meteorological conditions on the link quality of sensor networks.

The study presented in [4] supports greenhouse monitoring. It is based on the deployment of four sensors to monitor temperature, humidity, light, and the level of carbon Dioxide in a green house. The hardware platform sends compressed IPV6 packets over IEEE802.15.4 with a loss rate of 5%. Authors conclude through their experimental study that there is a relationship between the relative temperature and humidity inside a greenhouse such that the temperature increases once the humidity has decreased.

The focus of [5] is to understand signal propagation between fixed and a mobile sensor and evaluate the impact of meteorological conditions on signal propagation. Memsic wireless sensors with three integrated sensors( humidity, temperature, barometric pressure) have been used. Their study demonstrates that the signal variation at each position is small and the variations in the presence of large water bodies is influenced by changes in temperature, humidity, and the refractive index of the medium.

[6] Proposed an idea for power control in wireless sensor networks that depends on the evaluated temperature and humidity. They developed a controller that adapts power according to RSSI feedback.

The main research interest of [7] is to study the impact of meteorological conditions on 802.15.4 links. Authors of the paper evaluated different conditions in Sweden. Their research conclude that the temperature variations correlate the most with RSSI and PRR.

[8] Conducted experiments and designed a scheme that combines both temperature-aware link quality compensation and a closed-loop feedback process to adapt to link quality.

In this paper, we investigate how the temperature and humidity variations may affect link quality of XM1000 sensors. Our idea studies the influence of the two factors on RSSI and LQI and uses ANOVA (variance analysis) to recognize the importance of each parameter.

## 3. XM1000 SENSOR MOTE

The consolidation of micro-electro mechanical systems (MEMS) and digital wireless communications has led to the ability of deploying small devices that are application-specific. In particular, sensors are provided with sensing and communicating capabilities that overcome the limitations of traditional networks for remote real time monitoring. Many kinds of sensors are available in the market and are designed and manufactured by leader industrial companies: CrossBow, Atmel,… XM1000 mote module is based on TelosB specification which is the result of the research conducted in UC Berkely. TelosB is equipped by MSP430 microcontroller which is characterized by many features [9]:

- ✓ Complete system on chip that includes LCD control, ADC, I/O ports, RAM(10 KB), ROM, timer, UART,…
- ✓ Low power consumption: 4.2 nW/instruction.
- ✓ High speed: 300 ns/instruction a clock speed of 3.3MHZ.
- ✓ Risc set of 27 core instructions that make the process of programming very easy.
- ✓ 48 KB of flash memory.
- ✓ 802.15.4 compliant mode and consumes low power for communications.
- ✓ Current in active mode is 1.8 mA and the current in sleep mode is 5.1MA.

Three sensors are integrated on XM1000: temperature, humidity, and light sensors. Its dimensions are 81.9mm*32.5mm*6.55mm including USB connector. It is provided by 116KB of program flash, 8KB of data program and 1MB for logging. The radio frequency chip of XM1000 operates in 2.4-2.485 GH with a rate of 250Kbps and can cover a range of 120m outdoor and between 20 to 30 m indoor [10].

Some research works have used XM1000 for carrying experiments:

ManOs has been tested on XM1000 to evaluate the potential to make sensor programming accessible for a broader range of programmers [11].

[12] studied and compared a set of empirical models to estimate the path loss by means of XM1000.

[13] demonstrated that device heterogeneity may improve energy by moving resource intensive tasks to other nodes in the network. Their study has focused on imote2, Inga, Mulle v5-2, Sunspot, TelosB, and XM1000.

## 4. TinyOS

XM1000 is compatible with TinyOs2.x which is open project that was developed at University of California. It is based on Nesc language that is suited for programming embedded systems. Its syntax is similar to C syntax but NesC is enhanced to support concurrency and allows to structure a program into linked components to reduce the compile time. Unlike C, NesC locates all functions at compile time. NesC compilers generate a C file after processing the NesC file. The C file is then compiled to the desired microcontroller. Each NesC application program is divided into linked modules to avoid any explicit pointer. The configuration file specifies the links between the system modules however, the interface specifies the interaction between two components: the user which implements a set of events and the provides that implements a set of commands [14].

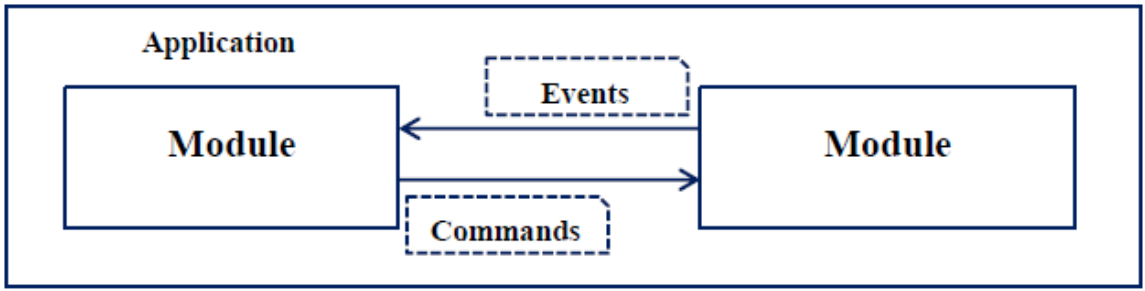

Figure1 . Component of NESC programming model.

The specification of each component (module or configuration) contains two parts: the signature and the implementation. The signature describes if the component use or provide any interface.

## 5. ANOVA

ANOVA is a statistical technique that takes one or more populations as input and tests the inequality among population means. More specifically, ANOVA analyses the effect of one or two qualitative factors upon a quantitative variable. ANOVA is based on hypothesis test [15]:

- $H0: \mu_1 = \mu_2 = \cdots \mu_k$ This means that all population means are equal. K is the total number of groups.

- $HA$: At least two of the means differ.

ANOVA calculates the F-test such that:

$$F = \frac{MSR}{MSE} \qquad (1)$$

$MSR$ is the mean square of the treatment and $MSE$ is the mean square of error.

$$MST = \frac{SST}{K-1} \qquad (2)$$

$$MSE = \frac{SSE}{n-K} \qquad (3)$$

$SST$ is the variation between however $SSE$ is the variation within and n is the number of samples in each group.

F is expected to be nearly equal to 1 if all population means are equals. Otherwise, the hypothesis $H0$ will be rejected.

## 6. EXPERIMENTAL STUDY

Our experimental study involves two XM1000 sensors. The first one is considered as a base station and sends periodically a message to the second one. Upon receiving a message, the target sensor measures the temperature, the humidity, RSSI, and LQI. The distance between the two sensors is around 20 Cm.

Figure 2 illustrates the hardware used in our experiment. The plotting of the variation of the measured RSSI according to the temperature is shown in figure 3. Linear regression has been applied and the color of the trendline is black. It is clear that there is a polynomial relation between RSSI and the temperature. Moreover, we can conclude from the result of the regression that RSSI is very high in the first group (12-27$^0$) and decreases once the temperature exceeds 27$^0$

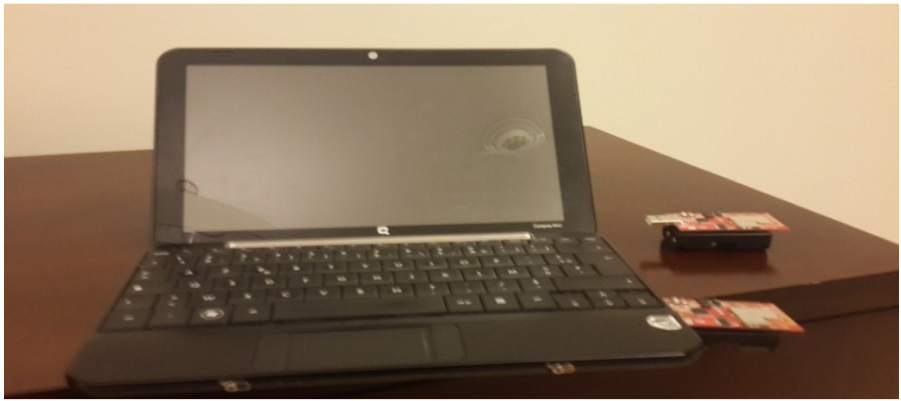

Figure 2. XM1000 sensors.

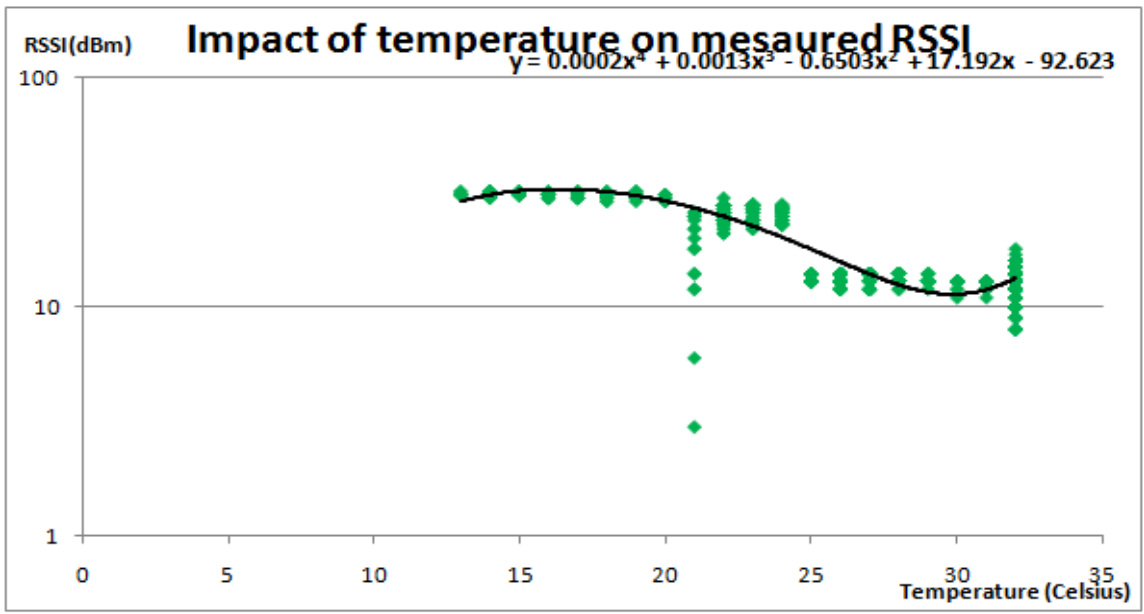

Figure 3. Temperature versus RSSI.

The plotting of the variation of the measured RSSI according to humidity is shown in figure 4. Linear regression has been applied and the color of the trendline is black. It is clear that there is a linear relation between RSSI and humidity. We can recognize two groups of RSSI measures after applying linear regression and RSSI increases in a linear manner from the first group that corresponds to humidity values ranging from 9 to 13% to the second one that corresponds to humidity values ranging from 13 to 15%.

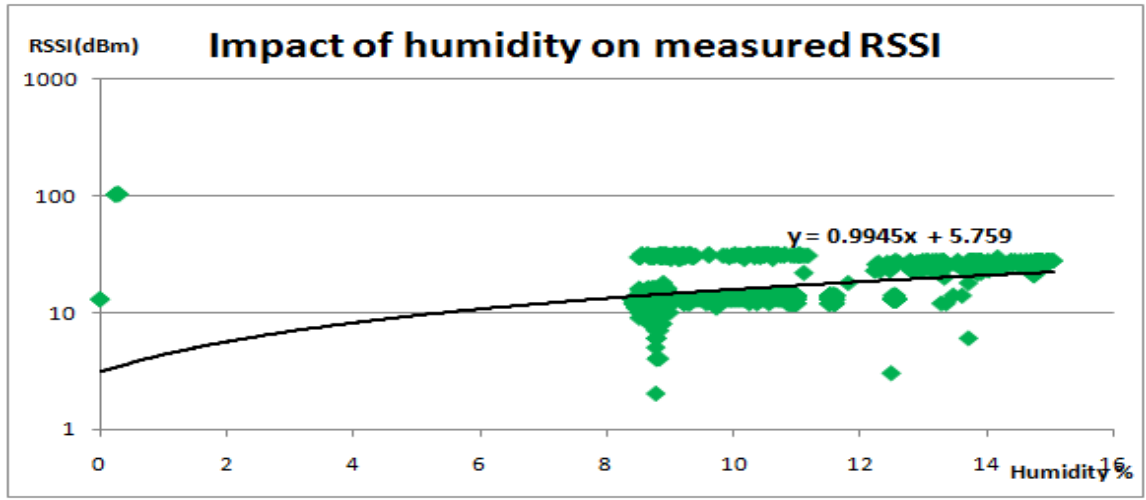

Figure 4. Humidity versus RSSI.

Figures 5 and 6 illustrate the variation of LQI according to temperature and humidity respectively. From the plots of linear regression, it is clear that the variation of LQI versus temperature and humidity is linear.

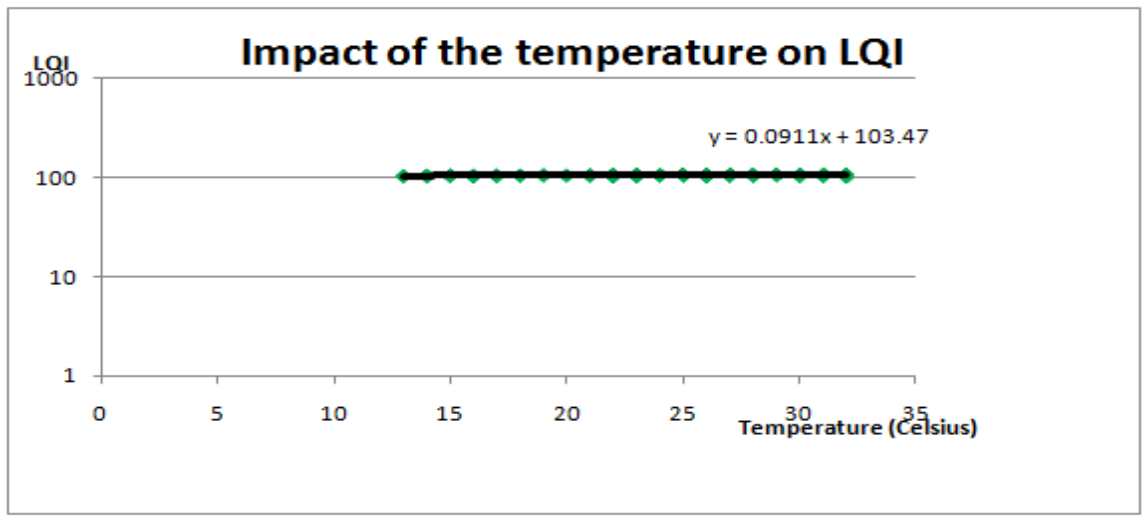

Figure 5. Temperature versus LQI.

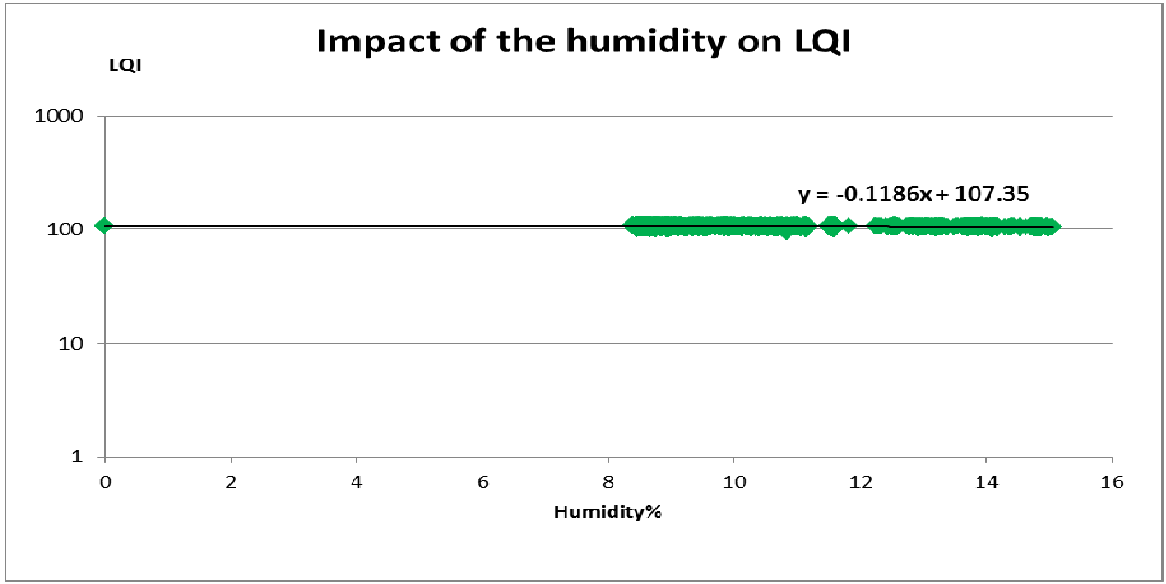

Figure 6. Humidity versus LQI.

We were asking the question about the important parameter that impacts RSSI as the function of variation of RSSI according to temperature is different from its function of variation according to humidity. To this end, we applied ANOVA after preparing three groups of temperature (high, warm, low) and two groups of humidity (high and low).

Three comments should be highlighted form table 1.

-The variance of each group is not far away from 1.

-The values for F are greater of F crit.

-P-values are smaller than 0.5.

So, H0 hypothesis is maintained and both temperature and humidity influence RSSI.

## 7. CONCLUSION

Estimating link quality is an important factor for many sensor protocols. Our experimental study concerned the estimation of both RSSI and LQI in a small XM1000 sensor network. Our results show that the impact of the temperature and humidity variations on LQI are linear. ANOVA analysis shows that RSSI is influenced by temperature as well as humidity. Hence it is necessary to save energy according to those two parameters. We will focus in the future on power control and saving in XM1000 sensor networks.

27

| Anova: Two-Factor With Replication | | | |
|---|---|---|---|
| SUMMARY | high humid | low humid | Total |
| *high temp* | | | |
| Count | 7 | 7 | 14 |
| Sum | 91 | 108 | 199 |
| Average | 13 | 15.42857143 | 14.21428571 |
| Variance | 0.666666667 | 7.952380952 | 5.565934066 |
| | | | |
| *warm* | | | |
| Count | 7 | 7 | 14 |
| Sum | 174 | 212 | 386 |
| Average | 24.85714286 | 30.28571429 | 27.57142857 |
| Variance | 5.80952381 | 1.238095238 | 11.18681319 |
| | | | |
| *cold* | | | |
| Count | 7 | 7 | 14 |
| Sum | 220 | 113 | 333 |
| Average | 31.42857143 | 16.14285714 | 23.78571429 |
| Variance | 0.285714286 | 57.47619048 | 89.56593407 |
| | | | |
| *Total* | | | |
| Count | 21 | 21 | |
| Sum | 485 | 433 | |
| Average | 23.0952381 | 20.61904762 | |
| Variance | 63.09047619 | 69.14761905 | |

| ANOVA | | | | | | |
|---|---|---|---|---|---|---|
| Source of Variation | SS | Df | MS | F | P-value | F crit |
| Sample | 1327 | 2 | 663.5 | 54.21595 | 1.38E-11 | 3.259446 |
| Columns | 64.38095238 | 1 | 64.38095238 | 5.2607 | 0.027756 | 4.113165 |
| Interaction | 877.1904762 | 2 | 438.5952381 | 35.83852 | 2.72E-09 | 3.259446 |
| Within | 440.5714286 | 36 | 12.23809524 | | | |
| Total | 2709.142857 | 41 | | | | |

Table 1. AVOVA results.

**Biography**

**Samia Allaoua Chelloug** received the Engineer degree (Computer Science) in 2003, Magister degree (Computer Science) in 2006, and the Phd degree (Networking) in 2013 (all from University of Constantine, Algeria).
From 2006 to 2013 she was an Assistant Professor in the Faculty of Computer Science in Constantine University.
In August 2013, she joined the Department of Networks and Communication Systems at Princess Nourah University (Riyadh, Kingdom of Saudi Arabia) where she is presently an Assistant Professor.
Her current research interests include: wireless sensor networks, cognitive radio, Vanets, and Fourth Generation communication standards.